\documentclass[twocolumn]{aastex631}
\usepackage{amsmath, esint}
\usepackage{subfigure}

\newcommand\spin{\ensuremath{a_*}}
\newcommand{\degree}{\ensuremath{^\circ}} 

\shorttitle{Accretion Flow Images}
\shortauthors{Papoutsis et al.}

\graphicspath{{./}{figures/}}

\begin{document}

\title{Jets and Rings in Images of Spinning Black Holes}

\author{Evan Papoutsis}
\affiliation{Department of Physics, University of Illinois, Urbana, IL 61801, USA}
\author{Michi Baub\"ock}
\affiliation{Department of Physics, University of Illinois, Urbana, IL 61801, USA}
\author{Dominic Chang}
\affiliation{Department of Physics, Harvard University, Cambridge, Massachusetts 02138, USA}
\affiliation{Black Hole Initiative at Harvard University, 20 Garden Street, Cambridge, MA 02138, USA}
\author{Charles F. Gammie}
\altaffiliation{also Department of Astronomy, Illinois Center for the Advanced Study of the Universe, and NCSA}
\affiliation{Department of Physics, University of Illinois, Urbana, IL 61801, USA}

\begin{abstract}

We develop a ``dual cone'' model for millimeter wavelength emission near a spinning black hole. The model consists of optically thin, luminous cones of emission, centered on the spin axis, which are meant to represent jet walls.  The resulting image is dominated by a thin ring. We first consider the effect of black hole's spin on the image, and show that the dominant effect is to displace the ring perpendicular to the projection of the spin axis on the sky by $2 \spin \sin i + \mathcal{O}(\spin^3)$. This effect is lower order in $\spin$ than changes in the shape and size of the photon ring itself, but is undetectable without a positional reference.  We then show that the centerline of the jet can provide a suitable reference: its location is exactly independent of spin if the observer is outside the cone, and nearly independent of spin if the observer is inside the cone.  If astrophysical uncertainties can be controlled for, then spin displacement is large enough to be detectable by future space VLBI missions.  Finally, we consider ring substructure in the dual cone model and show that features in total intensity are not universal and depend on the cone opening angle.

\end{abstract}

\section{Introduction} \label{sec:intro}

In 2019 the Event Horizon Telescope (EHT) Collaboration imaged the millimeter source at the center of the galaxy M87 \citep{EHT_I,EHT_III}.  The source, M87*, lies at the base of the first known extragalactic jet \citep{Curtis_1918}, and has an angular scale comparable to that expected for a black hole with mass inferred from stellar velocity distributions \citep{Gebhardt_2011}. 

The M87* image is a bright, asymmetric ring.  The ring is almost certainly produced by synchrotron emission from hot, magnetized gas close to a black hole.   Models had predicted the ring shape \citep{Falcke_2000, Broderick_2009, Dexter_2012, Mosc_2016}, produced by gravitational lensing of emission from close to and behind the hole.  The asymmetry is well explained by Doppler boosting \citep{EHT_V}.  The region on the observer plane where photon trajectories end on the horizon is commonly referred to as the {\em black hole shadow.}  The boundary of the shadow is called the {\em critical curve}.  The shadow overlaps, to within the modest precision of EHT, with the central brightness depression in observed images of M87*.

The critical curve depends on the dimensionless black hole spin $\spin$ and the inclination $i$ of the spin axis to the line of sight \citep{Bardeen_1973}.  A general parametric expression is available, but the effect of spin and inclination is most transparent if we expand the critical curve to order $\spin^2$ in sky angular coordinates $(x,y) c^2 D/(G M)$ which are (perpendicular, parallel) to the spin axis ($M \equiv$ black hole mass and $D \equiv$ distance to the source).\footnote{The coordinates $x,y$ are equivalent to Bardeen's (1973) $\alpha,\beta$.} Parameterizing the curve with angular coordinate $\tau$,
\begin{equation} \label{eq:x}
    x = 2 \spin \sin i + \sqrt{27}  (1 - \spin^2/18)  \cos\tau + \mathcal{O}(\spin^3),
\end{equation}
\begin{equation} \label{eq:y}
    y = \sqrt{27} (1 - \spin^2\cos^2 i/18) \sin\tau  + \mathcal{O}(\spin^3),
\end{equation}
\citep{GrallaLupsasca_2020b}.
The critical curve is circular to lowest order in $\spin$, the linear term shifts the shadow to the right ($+x$) when the projection of the spin axis on the sky is up ($+y$), and the quadratic term reduces the radius of the critical curve and makes it an ellipse.  The spin displacement could be a coordinate effect, and is meaningless unless endowed with physical significance.  In this paper we show that spin causes relative displacement of features on the image plane in a class of phenomenological ``dual cone'' models.

If the radiating plasma is optically thin then the intensity has an integrable singularity at the critical curve.  The intensity divergence is associated with photon trajectories that wrap around the hole many times and therefore have a long pathlength through the emitting plasma.  The divergence is an artifact of the optically thin approximation; at finite optical depth the intensity saturates (if the plasma is thermal then the saturation is at the Planck intensity) and the singularity is replaced by a finite intensity maximum.  

The image can be decomposed into a sequence of {\em subrings} indexed by $n = 0, 1, 2, 3, \ldots$ produced by photons that have circled the hole $n/2$ times \citep{Luminet, BHO}.  In equatorial disk models each subring produces a feature in the image plane and a corresponding interferometric signature in the uv plane.  Previous work has considered a variety of models for the source plasma, including thin equatorial disks \citep{Luminet}, spherically symmetric accretion \citep[e.g.][]{Narayan_2019}, and numerical (GRMHD) models \citep[e.g.][and many others]{Noble_2007}.  It is well known that in spherical accretion subrings do not produce image-plane features while equatorial disks do.  GRMHD models tend to produce complex structures near the critical curve, although there are signs of subrings in at least some models \citep{Johnson_2020}.  In this paper we ask whether in dual cone models the subrings produce image-plane features and whether they are universal or depend on model parameters.

This paper is organized as follows.  Section \ref{sec:pro} fixes notation for the Kerr metric and briefly describes how we calculate images from source models.  Section \ref{sec:model} describes the dual cone model.  Section \ref{sec:spin} discusses the signature of spin displacement in the dual cone model.  Section \ref{sec:photonring} discusses the structure of the dual cone model near the critical curve.  Section \ref{sec:conclusion} summarizes and presents our conclusions.

\section{Ray Tracing Method} \label{sec:pro}

In this paper we use the Kerr metric in Boyer-Lindquist coordinates $t, r, \theta,\phi$ with line element 
\begin{equation}
        \begin{split}
        ds^2=&-\left(1-2r/ \Sigma\right)dt^2-\left(4\spin r\sin^2\theta/\Sigma\right)dtd\phi\\
        &+\left(r^2+\spin^2+2\spin^2r\sin^2{\theta}/\Sigma \right)\sin^2\theta d\phi^2\\
        &+\left(\Sigma/\Delta\right)dr^2+\Sigma d\theta^2.
        \end{split}
\end{equation}
Here $\spin$ is dimensionless black hole spin and
\begin{subequations}
\begin{eqnarray}
    \Delta \equiv r^2 -2r + \spin^2,\\
    \Sigma \equiv r^2 + \spin^2 \cos^2\theta,\\
    A \equiv (r^2+\spin^2)^2-\spin^2\Delta. \sin^2\theta
\end{eqnarray}
\end{subequations}
Throughout we set $GM = c = 1$. 

For null geodesics with wave four-vector $k^\mu$ the integrals of the motion are
\begin{subequations}
\begin{eqnarray}
    k_t  = -E,\\
    k_\phi  = L_z, \\
    k_\mu k^\mu  = 0,\\
    k_\theta^2+\cos^2\theta(-\spin^2E^2+L_z^2/\sin^2\theta)  = Q,
\end{eqnarray}
\end{subequations}
where $E, L_z$, and $Q$ are the energy, angular momentum, and Carter constant.  The equations of motion are
\begin{subequations}\label{eq:motion}
\begin{eqnarray}
    \Sigma\frac{dr}{d\lambda}=\pm(R)^{1/2},\\
    \Sigma\frac{d\theta}{d\lambda}=\pm(\Theta)^{1/2},\\
    \Sigma\frac{d\phi}{d\lambda}=-(\spin E-L_z /\sin^2\theta)+\spin P/\Delta,\\
    \Sigma\frac{dt}{d\lambda}=-\spin(\spin E \sin^2\theta-L_z )+(r^2+\spin^2)P/\Delta,
\end{eqnarray}
\end{subequations}
where 
\begin{subequations}\label{eq:PRTdef}
\begin{eqnarray}
    P\equiv E(r^2+\spin^2)-L_z a,\\
    R\equiv P^2-\Delta[(L_z -\spin E)^2+Q],\\
    \Theta \equiv Q-\cos^2\theta [-\spin^2E^2+L_z ^2/\sin^2\theta].
\end{eqnarray}
\end{subequations}
\citep{Bardeen_1972}.
Notice that at turning points in $\theta$, $\Theta = 0$.

The intensity $I_\nu$ at a point on the sky is determined by the radiative transfer (Boltzmann) equation which, absent scattering, can be written as an ordinary differential equation along a geodesic parameterized by an affine parameter $\lambda$: 
\begin{equation}
    \frac{dI}{d\lambda} = -A I + J.
\end{equation}
Here $I \equiv I_\nu/\nu^3$, $A \equiv \nu \alpha_\nu$, and $J \equiv j_\nu/\nu^2$ are the invariant intensity, absorptivity, and emissivity, and $I_\nu$, $\alpha_\nu$, and $j_\nu$ are their frame-dependent counterparts, here evaluated in the plasma frame.  We assume the gas is optically thin and set $\alpha_\nu \rightarrow 0$.  Then 
\begin{equation}\label{eq:rtintegral}
    I = \int J d\lambda.
\end{equation}
The image $I_\nu(x,y)$ is sampled by evaluating (\ref{eq:rtintegral}) along  geodesics that end at the center of each pixel.  

\section{Dual Cone Model} \label{sec:model}

It is possible that some emission in black hole accretion flows arises from hot, thin layers along the walls of the jet (see Figure 4 of \citealt{EHT_V}; for earlier GRMHD models with emission in or near the jet, see \citealt{Dexter_2012, Mocibrodzka_2016}). To investigate the consequences of jet wall emission we have developed a toy model, which we will call the dual cone model.  

In the dual cone model emission arises near Boyer-Lindquist $\theta = \theta_c$ and $\theta = \pi - \theta_c$:
\begin{equation}
    j_\nu(r,\theta;\theta_c) = C r^{-5}\left(\mathrm{e}^{-\left(\frac{\theta-\theta_c}{w}\right)^2}+\mathrm{e}^{-\left(\frac{\theta-(\pi-\theta_c)}{w}\right)^2}\right).
\end{equation}
Here $C$ is an arbitrary constant.  We set $w = 1/128$ so that the emissivity is approximately a $\delta$ function in $\theta$ around $\theta_c$.  Notice that $j_\nu$ is independent of $\nu$, which is appropriate close to the synchrotron peak.  

The invariant emissivity $J = j_\nu/\nu^2$ depends on $\nu = - k_\mu u^\mu$, where $u^\mu$ is the plasma four-velocity.  We adopt a basic infall model for simplicity, with $u^\theta = 0$, angular momentum $u_\phi = 0$, and $-u_t = 1$.  Then $u^r$ is determined from $u_\mu u^\mu = -1$.  Different inflow and outflow models were tested, and although the flow model affects the position of the brightness maxima, it does not affect the location of the critical curve and the jet boundaries on the image plane. 

We have made several arbitrary choices in assigning the emissivity and four-velocity.  Our dual cone model is just one of a family of models with varying emissivity profiles, velocity profiles, angular momenta, etc.  Since the dual-cone model is purely illustrative we will retain only a single parameter, the cone opening angle $\theta_c$. 

\section{Spin Measurement} \label{sec:spin}

Extragalactic jets are likely powered by the \cite{1977MNRAS.179..433B}, or BZ, effect.  The BZ model taps the energy that is stored, flywheel-like, in spinning black holes.  The energy is extracted by magnetic fields that are frame-dragged by the hole.  It is thus a prediction of the BZ model that the black hole in M87* should be spinning.  Can millimeter VLBI be used to directly measure M87*'s spin?

VLBI spin measurements have been of interest for some time.  Methods have been proposed that use the ring shape, subrings, or lensing time delays \citep[e.g.][]{Falcke_2000, Takahashi_2004, Broderick_2005,  Bambi_2009, Hioki_2009, Moriyama_2015, Younsi_2016, Saida_2017, Wei_2019,  Dokuchaev_2019, Moriyama_2019, Thompson_2019, Johnson_2020, Gralla_2020, Hadar_2021, Wong_2021b, Broderick_2022}.  A method that relies on the structure of emission inside the critical curve has been proposed by \citet{Chael_2021} for equatorial disk emission models.  Here we propose a generalization of Chael et al.'s technique (see also \citealt{Takahashi_2004, Dokuchaev_2019}) that we motivate using the dual cone model.

First, notice that the leading order in $\spin$ correction to the critical curve is the $\mathcal{O}(\spin)$ displacement perpendicular to the projection of the spin axis on the sky.  The magnitude of this correction to the center of the critical curve is larger than the correction to the diameter when $\spin < 2 \sqrt{3} \sin i$. The correction to the center is therefore larger at $i > \arcsin(\spin/(2 \sqrt{3})) \approx 16.5 \spin$\degree.

The spin correction to the ring diameter is obviously accessible to interferometric measurement, although it is always smaller than $2 \sqrt{27} - 9 \approx 1.39$ (for M87$^*$, about $5\mu$as), which is reached for $\spin = 1, i = 90$\degree. For M87*, with $i \approx 20$\degree, the effect is smaller and always less than $0.82$.  For $(\spin,i) = (0.5,20 \degree)$ the correction to the (diameter, center) is $(0.14, 0.34)$; for $(\spin,i) = (1, 20 \degree)$ it is $(0.58, 0.68)$. 

The spin correction to the ring center is inaccessible to interferometric measurement, however, unless there is a reference feature to which it can be compared \citep{Takahashi_2004}.  For the dual cone model, a reference is provided by the centerline of the jet.

The simple geometry of the dual cone model permits the jet boundaries to be projected analytically onto the observer ($x,y$) plane if the observer lies outside the cone.  A point on the image is bright where the corresponding geodesic is tangent to the cone at $\theta = \theta_c$ and therefore has a long path length through the emitting region.  The geodesic has a tangent point on the cone if its $\theta$ turning point $\theta_t$ coincides with the cone opening angle $\theta_c$.  We will refer to the locus of points on the image plane where $\theta_t = \theta_c$ as the cone boundary.

From Equations (\ref{eq:motion}b) and (\ref{eq:PRTdef}b) the turning points in $\theta$ for geodesic motion lie at 
\begin{align}\label{eq:Theta0}
    \Theta(\theta_t)/E^2 = 0,
\end{align}
with roots parameterized by the energy-normalized integrals of motion $p \equiv L_z/E$ and $q^2 \equiv Q/E^2$. 
The screen coordinates of the geodesic are 
\begin{eqnarray}
    x =-\frac{p}{\sin i}\label{eq:obsxdef}\\
    y =\sqrt{q^2+(\spin^2-x^2)\cos^2 i}\label{eq:obsydef}
\end{eqnarray}
\citep{Bardeen_1973}.  Notice that this solution includes geodesics that are inside the horizon and do not reach the observer.  Plugging Eqs. \!\eqref{eq:obsxdef} \!and \!\eqref{eq:obsydef} \!into Eq. \!\eqref{eq:Theta0},
\begin{equation}
    \begin{split}
    0
    =& y^2 + (x^2 - \spin^2) \cos^2 i + \spin^2 \cos^2\theta_t\\
    &- x^2 \cot^2\theta_t \sin^2 i
    \end{split}
\end{equation}
and the turning points appear on the observer screen at
\begin{equation}\label{eq:xsqsol}
    x^2 = \left(\spin^2 + \frac{y^2}{\cos^2\theta_t - \cos^2 i}\right) \sin^2 \theta_t.
\end{equation}
which is even in $x$.  

Solutions to Equation (\ref{eq:xsqsol}) contain the locus of points on the image plane whose corresponding geodesics have turning points at $\theta_t$.  Setting $\theta_t=\theta_c$, Equation (\ref{eq:xsqsol}) defines conic sections on the image plane that are dependent on observer inclination $i$.  For ``inside-the-cone'' observers ($\cos^2\theta_c<\cos^2i$) the solutions form an ellipse, and for ``outside-the-cone'' observers  ($\cos^2\theta_c>\cos^2i$) the solutions form a hyperbola.  

A geodesic's trajectories in radius and latitude ($r,\theta$) are determined separately by independent equations (see \ref{eq:motion}).  As a result the $\theta$ turning point need not be reached. Points on the conic sections can be bright only if the turning point {\em is} reached.  We have numerically explored the relationship between the critical curve, the cone boundary, and the ``cone base'', the projection on the sky of the cone-horizon intersection\footnote{We use ``projection'' to mean the lowest order projection, since there are infinitely many additional images of the cone-horizon intersection formed near the critical curve.}.

\begin{figure}
    \includegraphics[width=\linewidth]{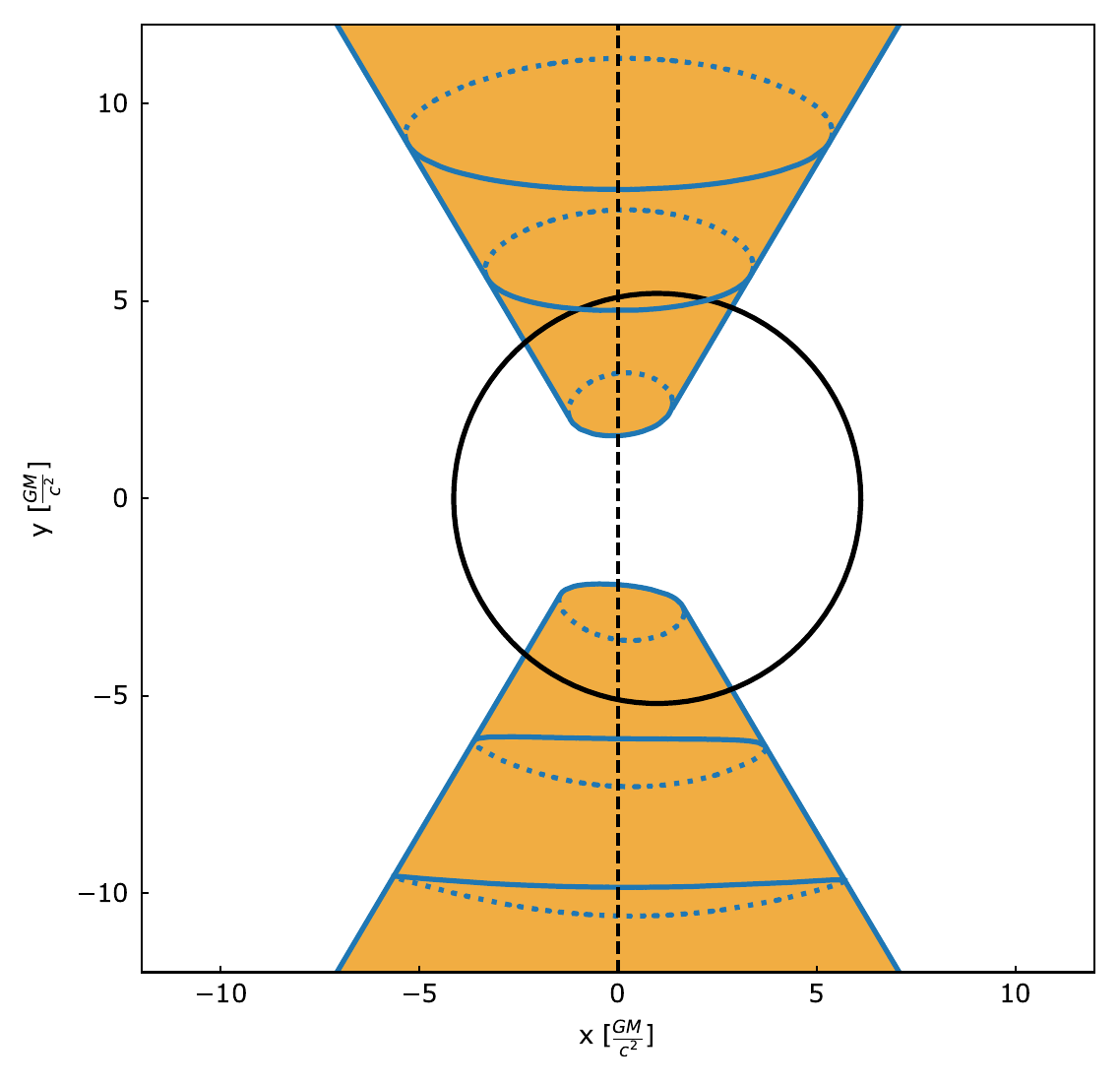}
    \caption{Map of the image plane, with geodesics that intersect the dual cones in orange.  The model has $\theta_c$ = 30\degree, $i$ = 80\degree, and $\spin$ = 0.5.  The midline of the cones is shown as a vertical black dashed line, and the critical curve as a rightward-displaced ellipse.  The orange region is bounded at the left and right by a hyperbola given by the solution to Equation (\autoref{eq:xsqsol}), and otherwise by the cone base, where the cone intersects the horizon.  The remaining blue solid and dashed curves show the projection of sections through the cone at $r = 6$ and $r = 10$.}
    \label{fig:jet_cone_shift}
\end{figure}

Figure~\ref{fig:jet_cone_shift} shows the image plane for $\theta_c = 30^\circ$, $i = 80^\circ$, and $\spin = 0.5$, which is an outside-the-cone case.  Points within the orange region reach the cones and those outside do not. The boundaries of the orange region are formed by the solution to Equation (\ref{eq:xsqsol}), found analytically, and by the cone base, found numerically.

Figure~\ref{fig:jet_cone_shift} also shows the critical curve as a solid curve that is clearly offset from the centerline of the cone.  The projection of the intersection of the $r=6$ and $r=10$ surface with the cone were found numerically and appear on the plot as a combined solid line, for the side of the cone closest to the observer, and dashed line, for the side of the cone furthest from the observer.

\begin{figure*}
        \fig{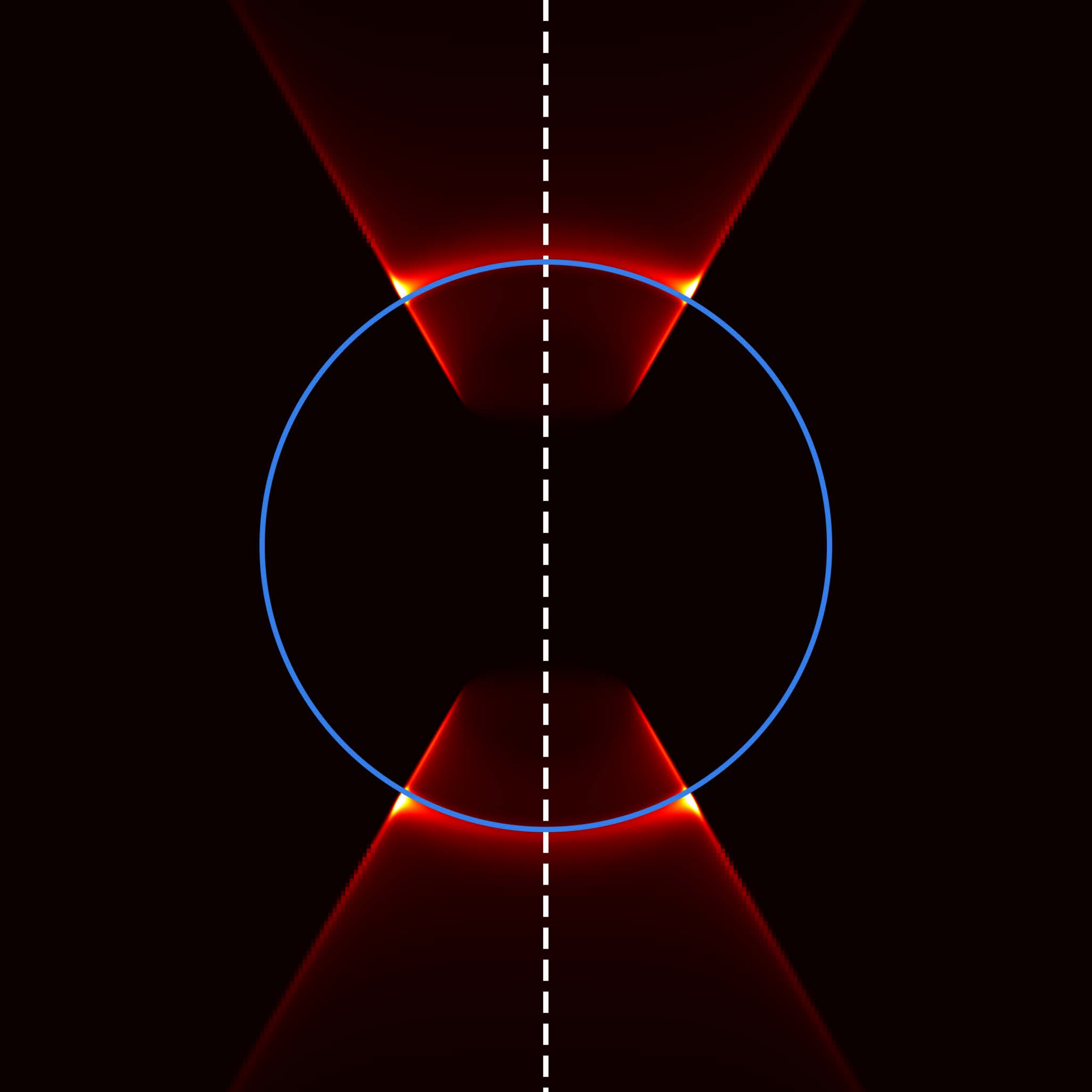}{0.24\textwidth}{}
        \fig{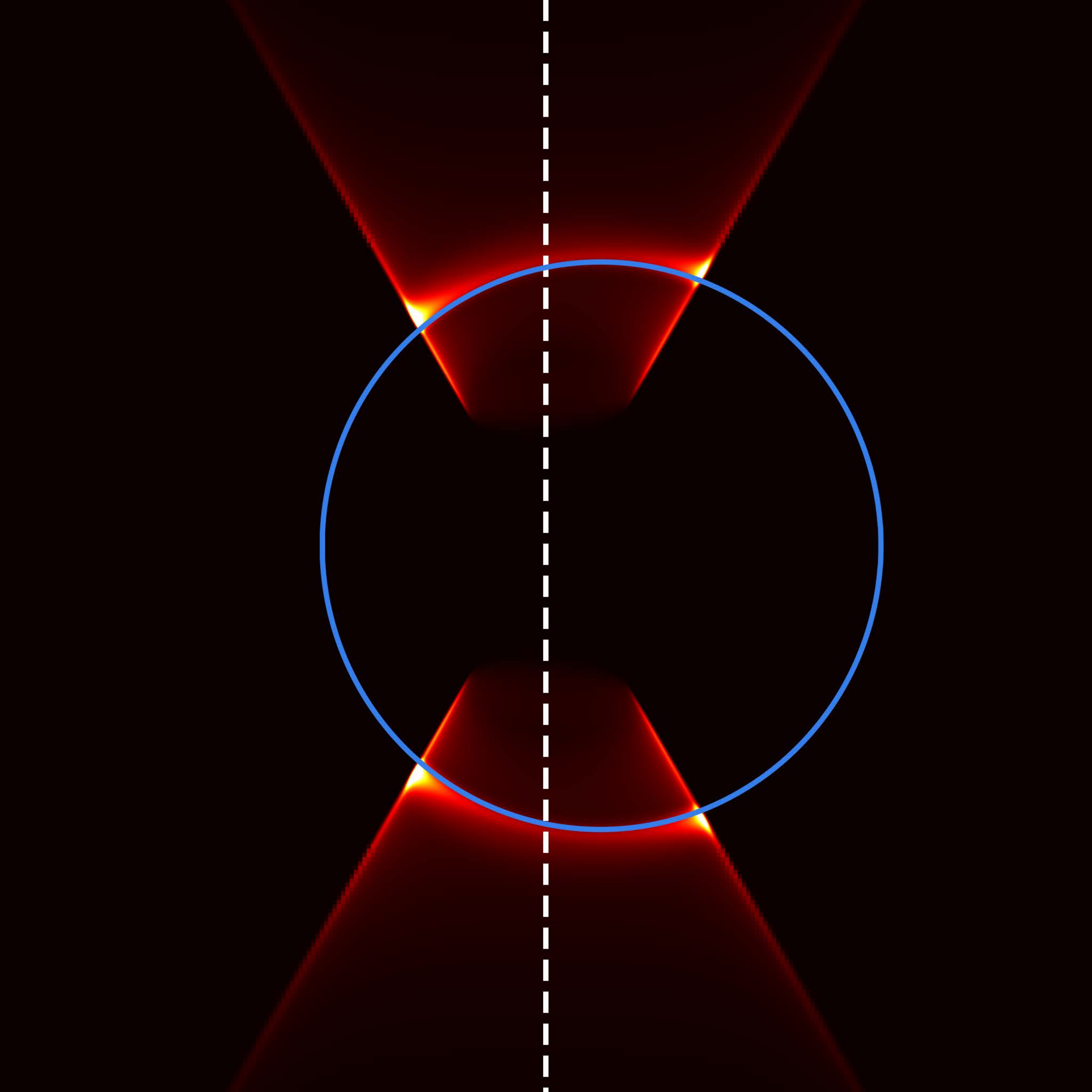}{0.24\textwidth}{}
        \fig{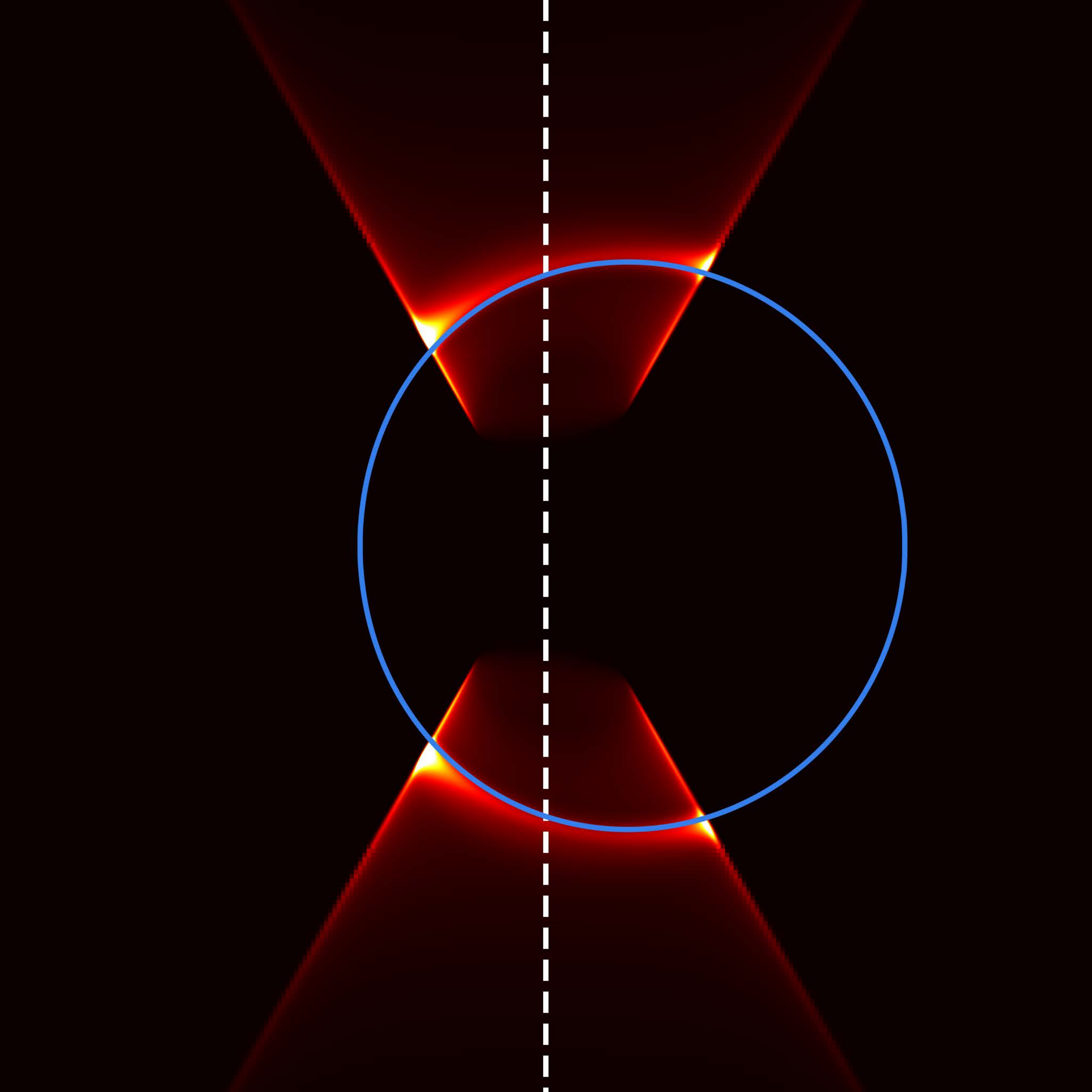}{0.24\textwidth}{}
        \fig{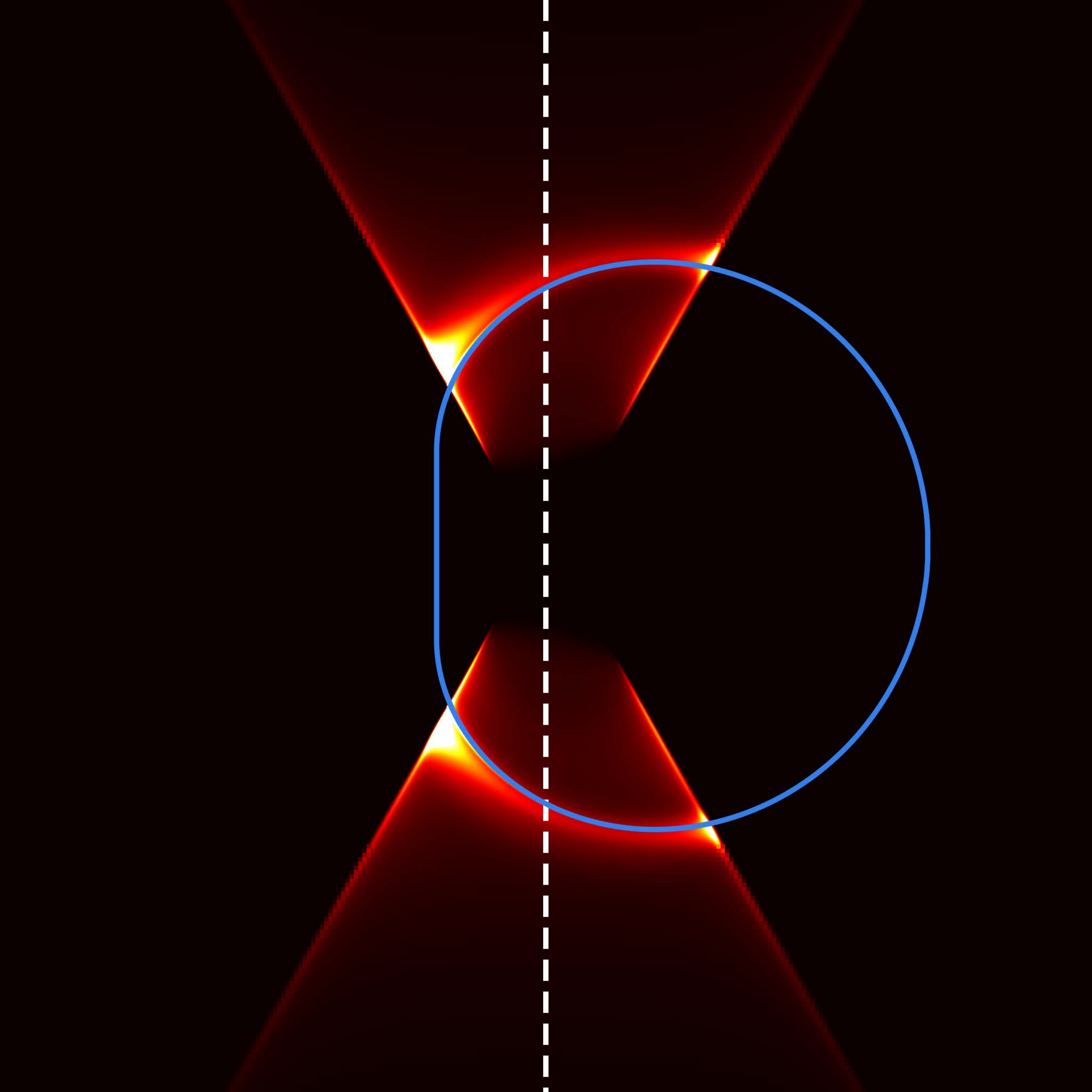}{0.24\textwidth}{}
        \fig{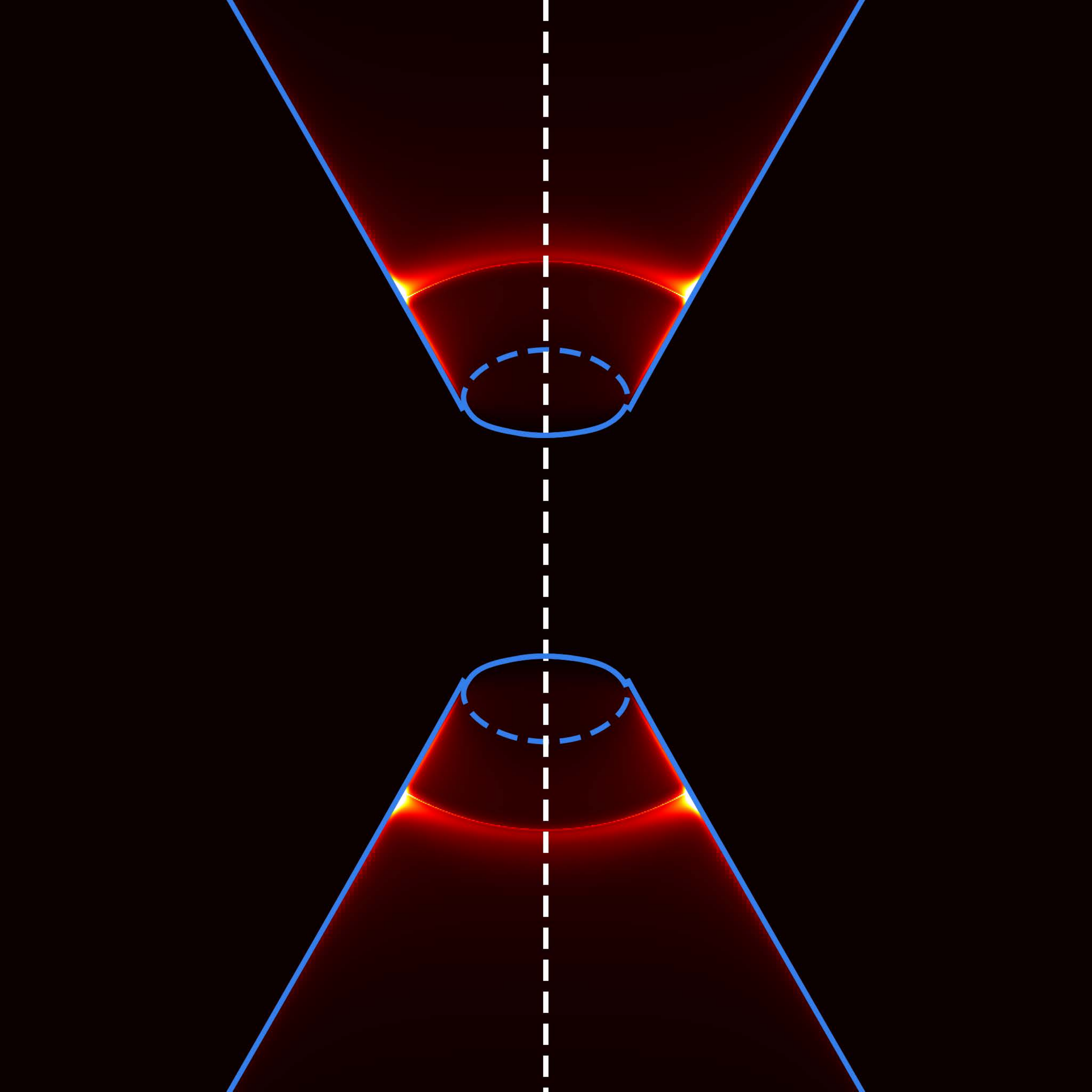}{0.24\textwidth}{(a) $\spin$=0.0}
        \fig{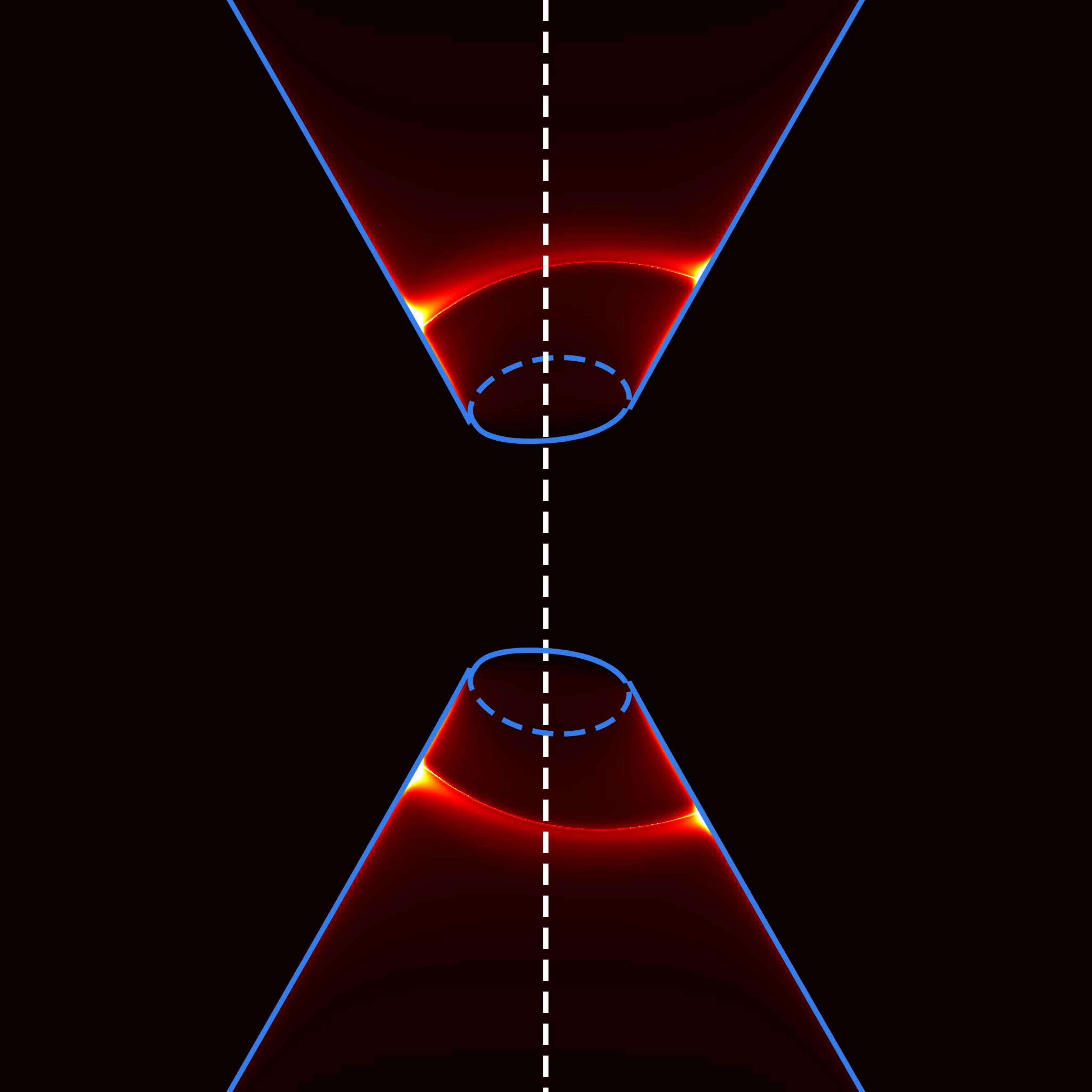}{0.24\textwidth}{(b) $\spin$=0.5}
        \fig{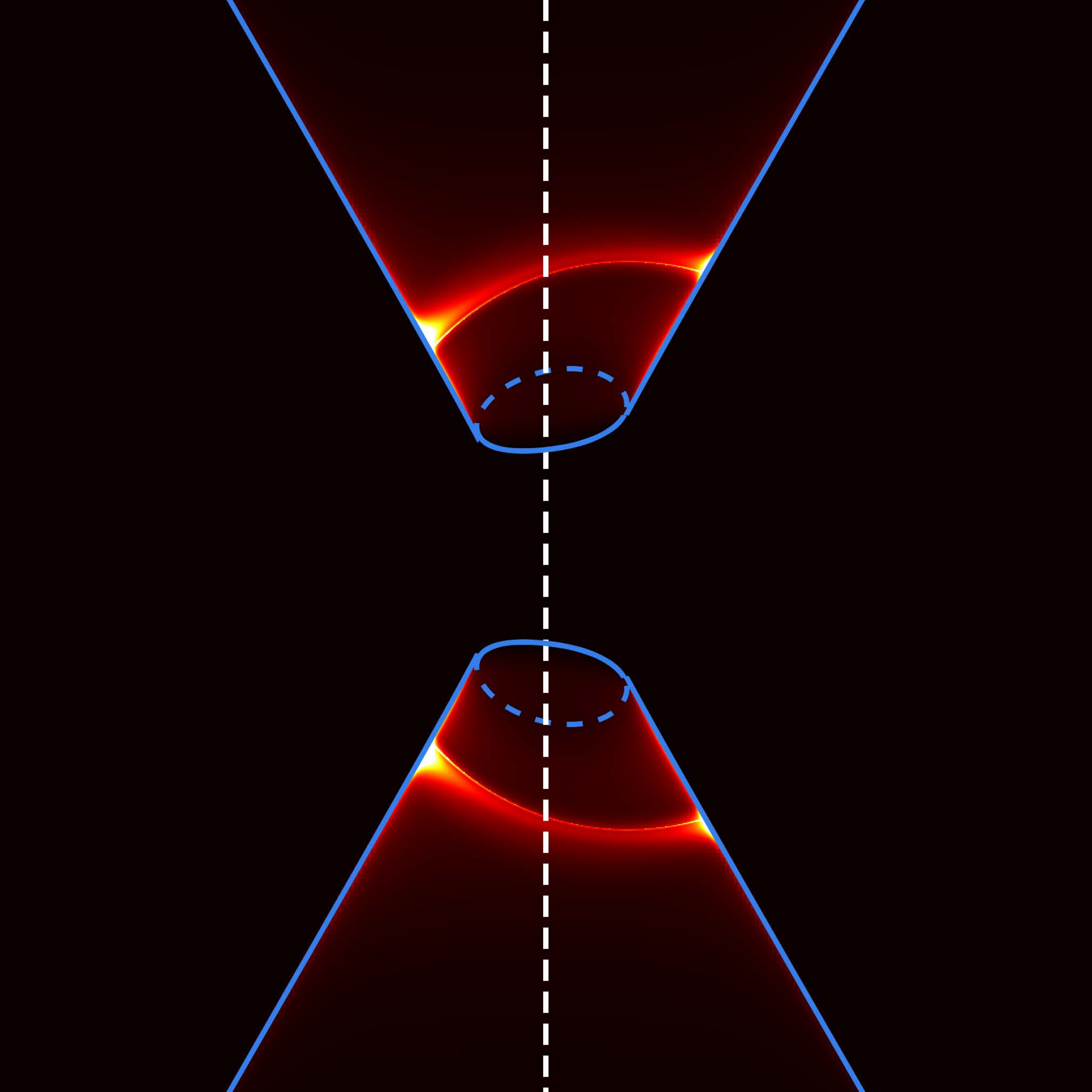}{0.24\textwidth}{(d) $\spin$=0.75}
        \fig{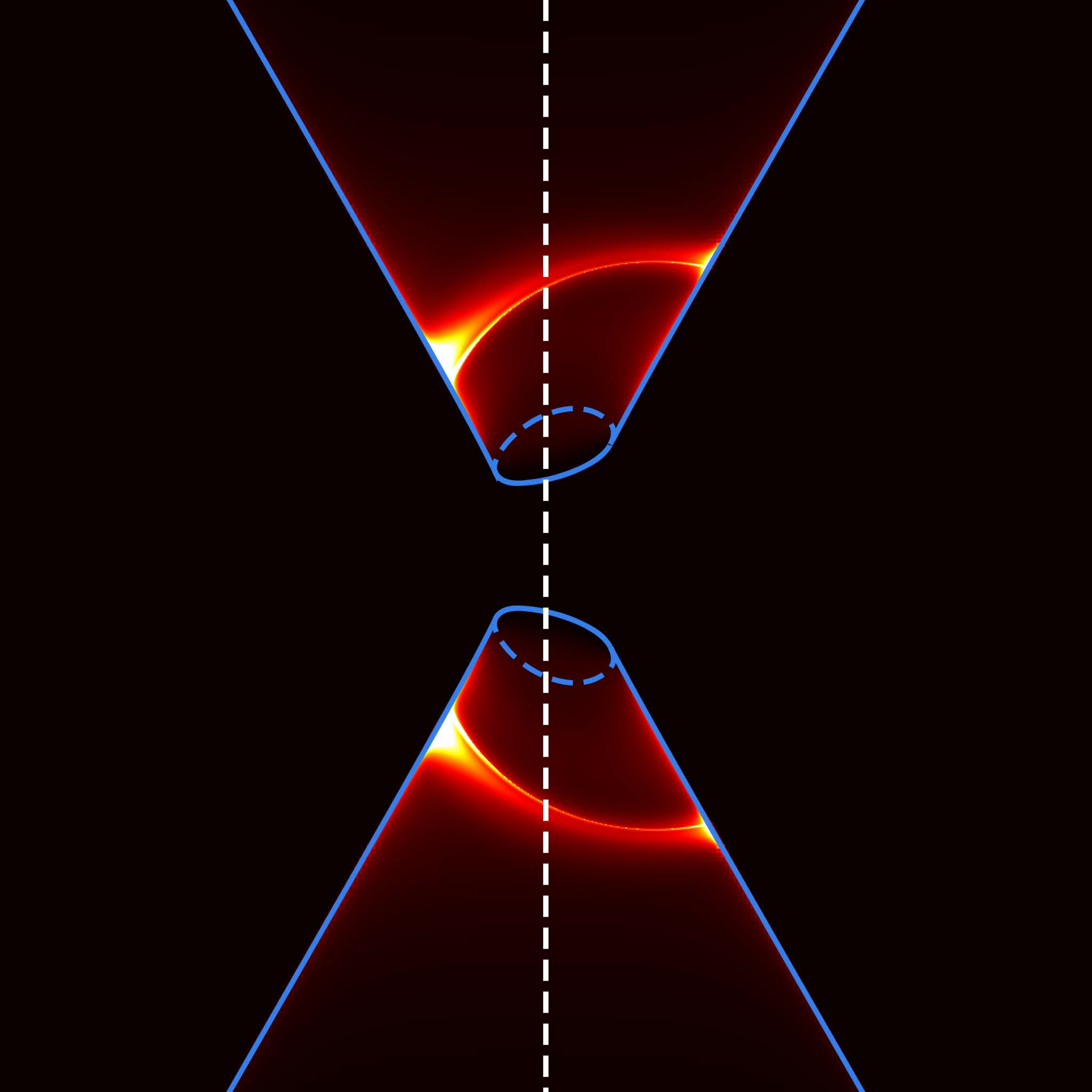}{0.24\textwidth}{(e) $\spin$=1.0}
        
\caption{``Outside-the-cone'' images of the $\theta_c = 30\degree$, $i = 90\degree$ dual cone model at $\spin = 0, 0.5, 0.75$, and $1.0$. The top row overlays the analytic critical curve in blue and the bottom row overlays the cone boundaries.
\label{fig:i_90_spin}}
\end{figure*}

\begin{figure*}
        \fig{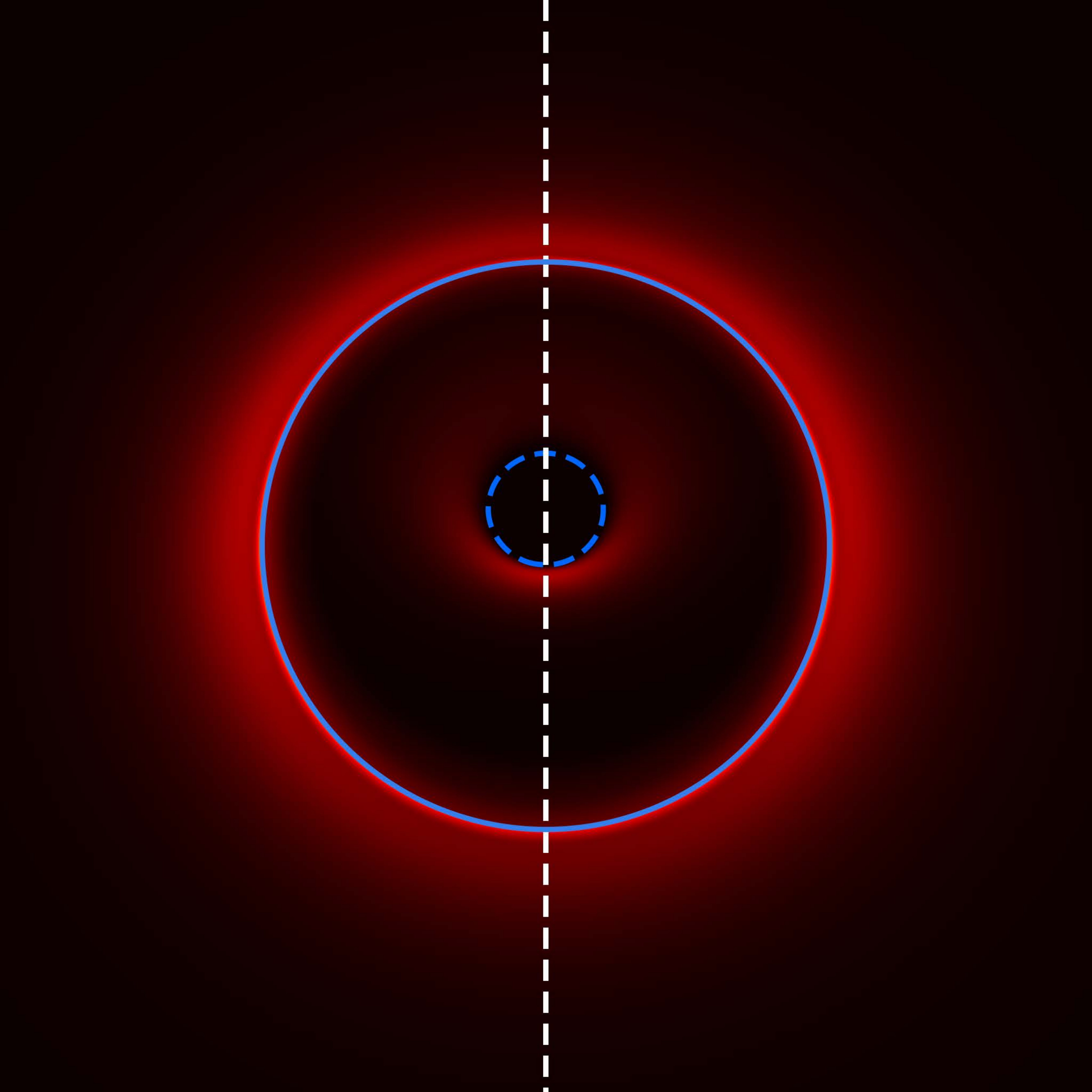}{0.24\textwidth}{(a) $\spin$=0.0}
        \fig{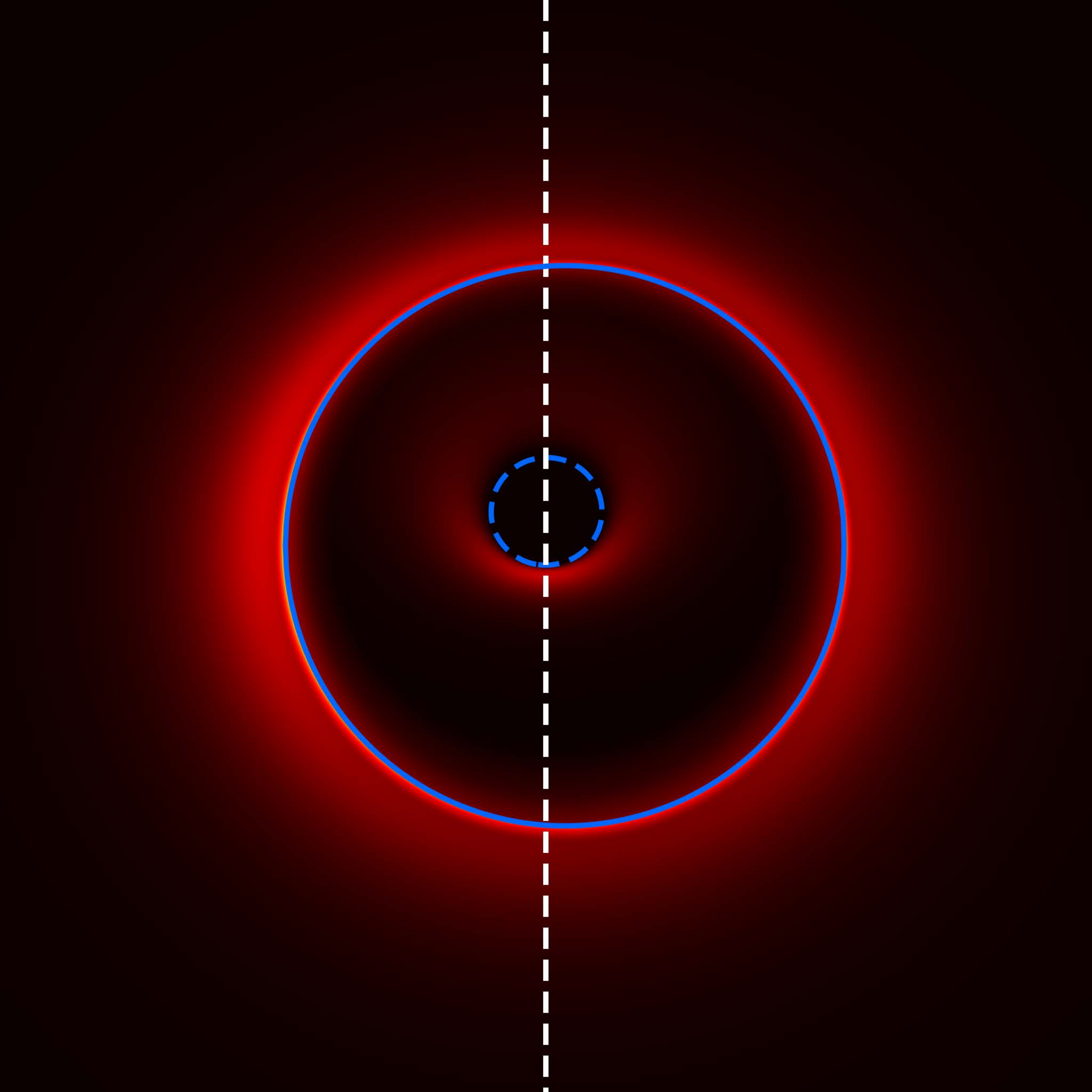}{0.24\textwidth}{(b) $\spin$=0.5}
        \fig{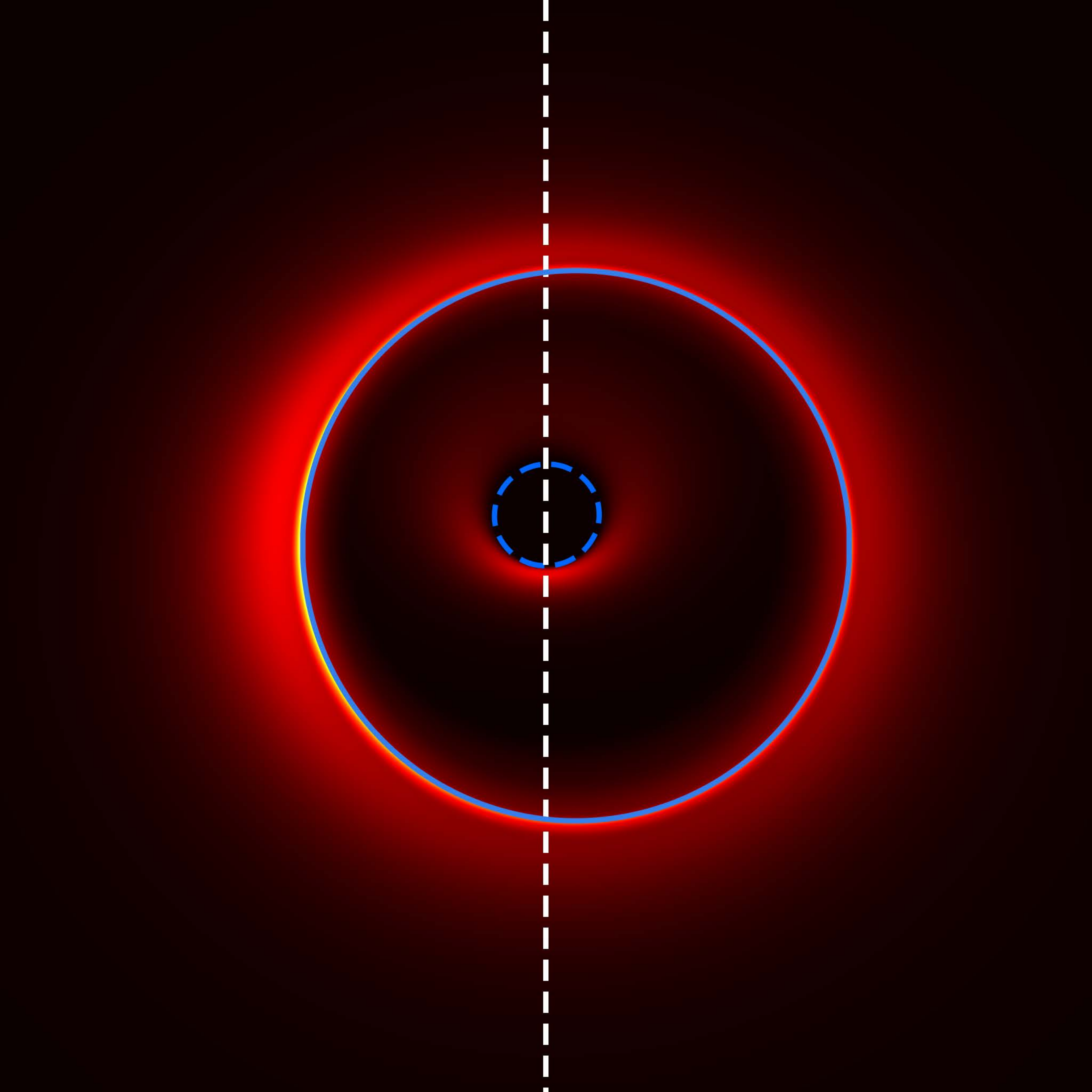}{0.24\textwidth}{(d) $\spin$=0.75}
        \fig{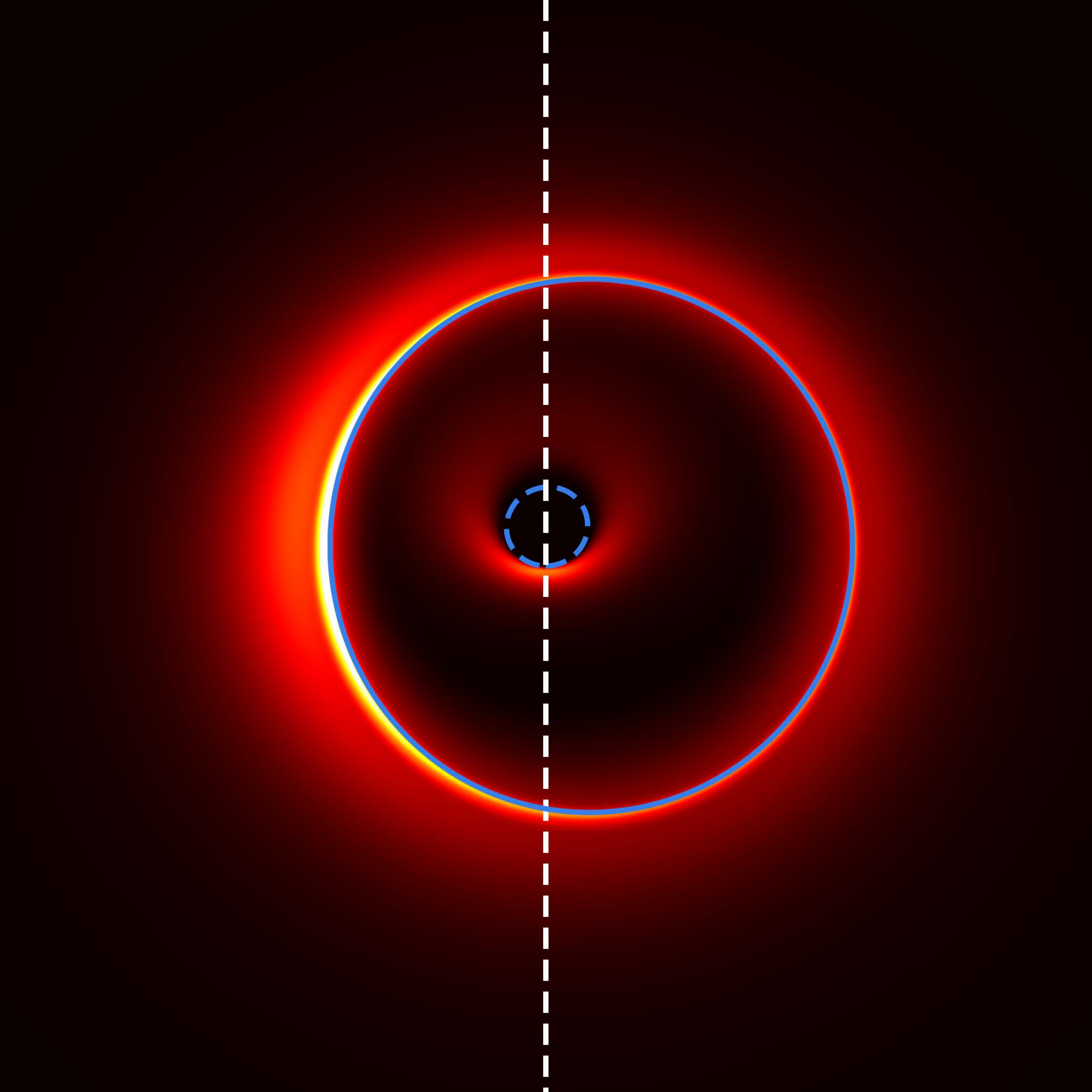}{0.24\textwidth}{(e) $\spin$=1.0}
\caption{``Inside-the-cone'' images of the $\theta_c = 30\degree$, $i = 20\degree$ dual cone model at $\spin = 0, 0.5, 0.75$, and $1.0$. The cone base (dashed) and critical curve (solid) are outlined in blue.
 \label{fig:i_20_spin}}
\end{figure*}
Figure \ref{fig:i_90_spin} shows a set of sample images at spins 0, 0.5, 0.75, and 1 for an observer with $i = 90$.  The top row contains a blue line showing the critical curve location; the bottom row shows the cone boundary (solid) and the cone base (dashed).  All images show the cone centerline as a dashed white line.  Evidently the cone boundaries and the critical curve provide an excellent guide to the location of bright features on the image for outside-the-cone observers.  Notice again that the critical curve is displaced to the right with respect to the cone centerline as spin increases.  

The situation for inside-the-cone observers is different and illustrated in Figure \ref{fig:i_20_spin}, which shows a set of sample images at spins 0, 0.5, 0.75, and 1 for an observer with $i = 20$\degree.  Again the blue curve shows the critical curve, a dashed blue curve shows the cone base, and a dashed white line shows the cone centerline.  Now only two structures are visible: the bright region surrounding the critical curve, and a secondary ring interior to the critical curve and surrounding the cone base. 

For inside-the-cone observers the solutions to Equation (\ref{eq:xsqsol}) form an ellipse with semi-major axis $= G M/c^2$ aligned normal to the cone centerline. For most parameters we have studied numerically the entire ellipse is interior to the cone base.  The turning points are not reached and there is no corresponding feature in the image plane.  For $i$ close to $\theta_c$ the turning point solutions sit just above the bottom edge of the secondary ring and form a bright feature.  In general the turning point solutions for inside-the-cone observers do not provide a guide to bright features on the image plane.  

The secondary ring's location is almost independent of spin. It experiences a slight asymmetrical deformation, with larger deformation at larger $\theta_c$, $i$, and \spin.  For $\theta_c < 60\degree$, the secondary ring displacement is always less than 17\% of the critical curve displacement.  For $\theta_c > 60\degree$, the secondary ring  displacement is less than 40\% of the critical curve's (see \citealt{Chael_2021} for an example).  Figure \ref{fig:i_20_spin}, where $i = 20$\degree and $\theta_c = 30$\degree, shows that the displacement between the secondary ring and the critical curve increases as spin increases.  Evidently the secondary ring's centerline can still serve as an approximate reference for displacement of the critical curve.

If it were possible to image M87* directly at 230GHz with lunar-orbit-length  baselines ($\approx 6 \times 10^{11}$G$\lambda \approx 0.3\mu as$ resolution) then ring displacement would be resolvable if $a \gtrsim 0.13$.  It is more likely, however, that the available data would take the form of visibility amplitudes measured along, for example, the 14 day ($\approx 40 G M/c^3$) eccentric orbit that is current proposed for the MUVE space-VLBI instrument on SALTUS.  Extraction of a spin displacement signal from this data is beyond the scope of this paper and is better done with more realistic models.  

One concern for future measurements is turbulent fluctuations in the source that change the image on short timescales.  The fluctuations might be confused with spin displacement.  Time-averaging may eliminate this problem if the fluctuations are well behaved, the observing cadence is high enough, and the mission duration is long enough.  Notice that the inner ring displacement is already visible (in the direction perpendicular to the projection of the spin vector on the sky) in the time-averaged images of GRMHD models in Figure 2 and 3 of the EHTC's M87 Paper V \citep{EHT_V}. 

\section{Application to Ring Substructure} \label{sec:photonring}

Substructure in the ring (image-plane features associated with the $n = 1$ and $n = 2$ subrings) can be used to measure spin \citep{Broderick_2022} for accretion models in which emission is limited to the midplane.  The question arises whether the subring structure is flow dependent. Strictly speaking the answer to this question is already known: in spherical accretion the intensity varies smoothly with impact parameter and ring substructure is absent \citep{Narayan_2019}. Since intensity features are associated with crossing planes of emission it seems likely that the dual cone model also produces flow-dependent image-plane structure.  Here we consider how dual-cone ring substructure depends on $\theta_c$. 

Null geodesics that lie sufficiently close to the critical curve on the image plane have emerged from close to the photon orbit, and have therefore been deflected by $n > 1$ half-rotations.  For $|b - b_c| \ll 1$ the deflection angle is proportional to $\mathrm{log}(b-b_c)$ \citep{Luminet}.  In this regime a ray that is deflected by $\Delta\phi \gg 1$ should be almost identical to a ray that is deflected by $\Delta\phi+2n \pi$. 

\begin{figure}
    \includegraphics[width=\linewidth]{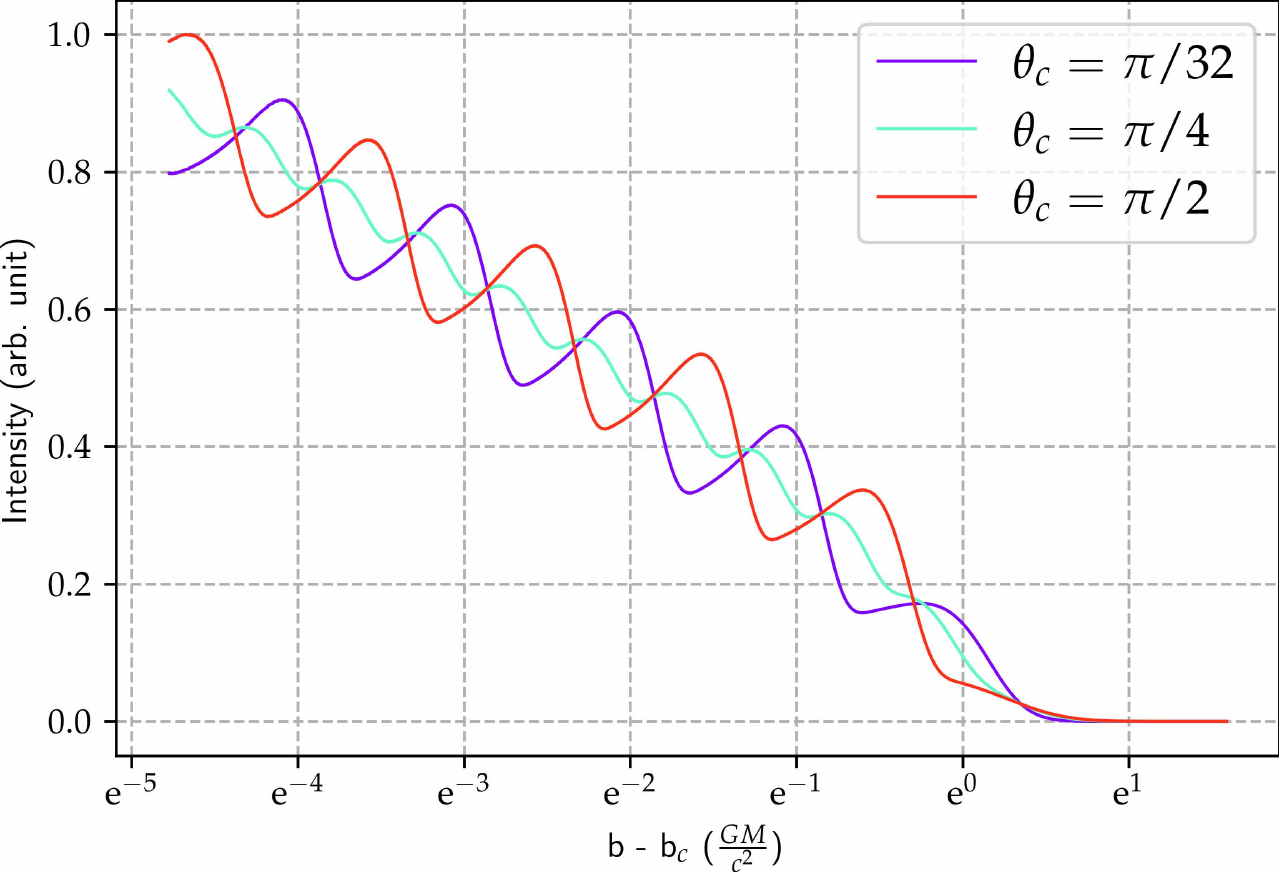}
    \caption{Intensity as a function of impact parameter for dual cones with opening angles $\theta_c = \pi/32, \pi/4, \text{ and }\pi/2$}
    \label{fig:coneimage}
\end{figure}

Figure \ref{fig:coneimage} shows $I_\nu(b; \theta_c)$ for $\theta_c = \pi/32, \pi/4, \text{ and }\pi/2$ with a simple face-on ($i = 0$) model with $\spin=0$, so that the critical curve lies at impact parameter $b^2 = x^2 + y^2 = 27$. Notice that the location and amplitude of peaks in $I_\nu(b)$ vary with $\theta_c$. Ring width and diameter is increased by a factor of $\sqrt{e}$ from the equatorial plane model to the low opening angle dual cone model.  At certain $\theta_c$, e.g. $\theta_c = \pi/4$, the interval between peaks is halved and there are two maxima associated with each subring (maxima arising from the ray crossing the $\theta=\pi/4$ and $\theta=3\pi/4$ planes).
While each pass through the emission region correlates with an intensity increase, whether this increase produces a detectable maximum is dependent on $\theta_c$ and $i$. 

\begin{figure*}
    \fig{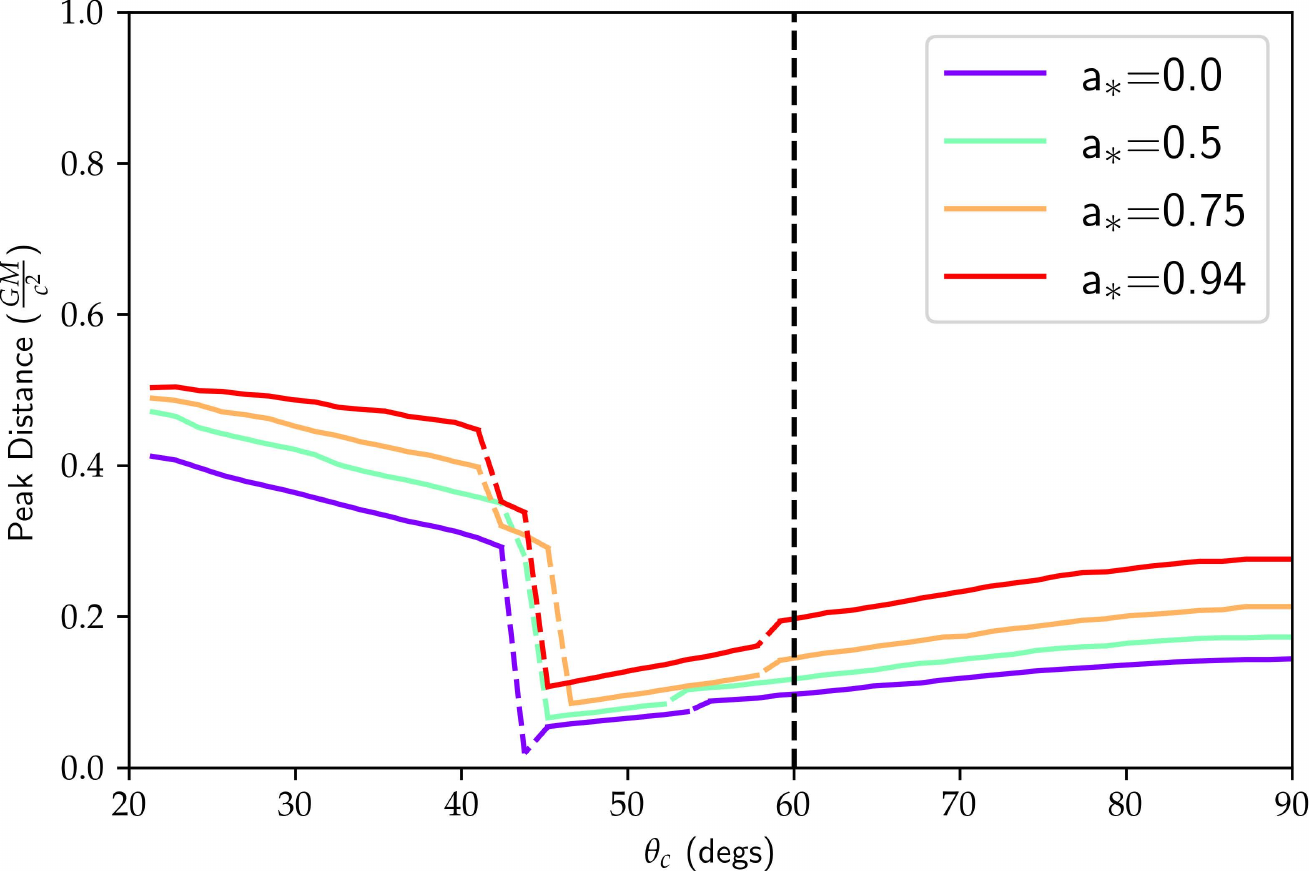}{.48\textwidth}{}
    \fig{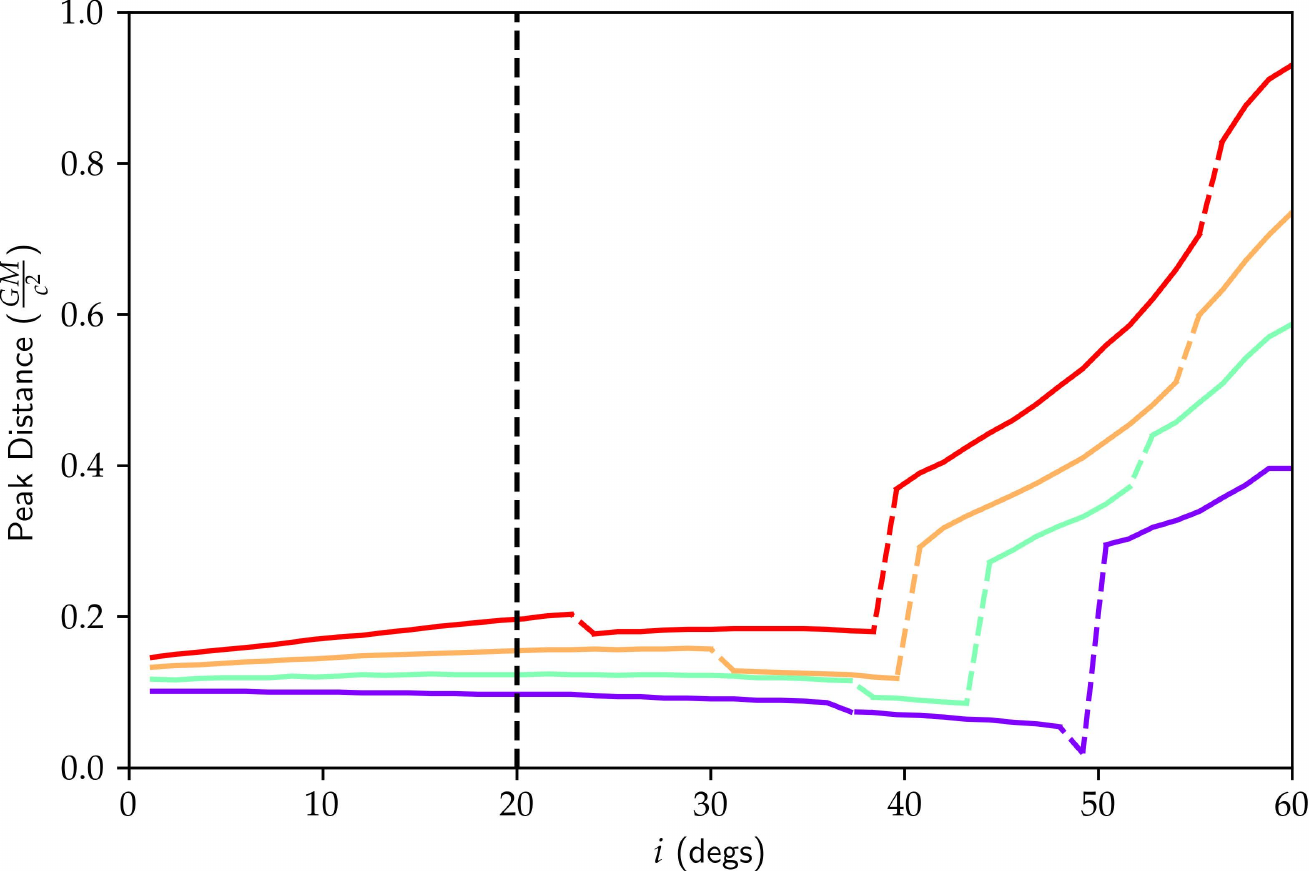}{.48\textwidth}{}
    \fig{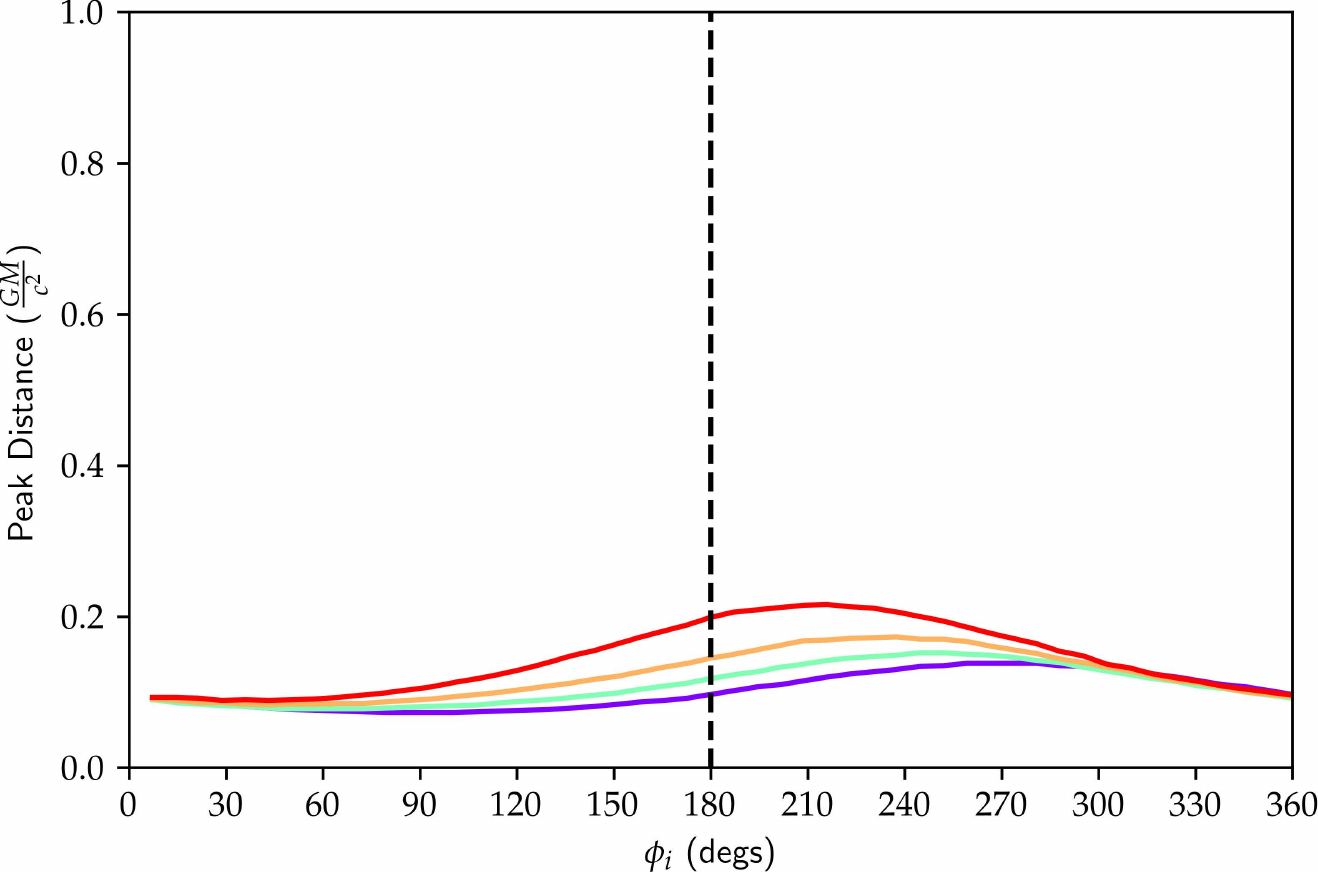}{.48\textwidth}{}
    \fig{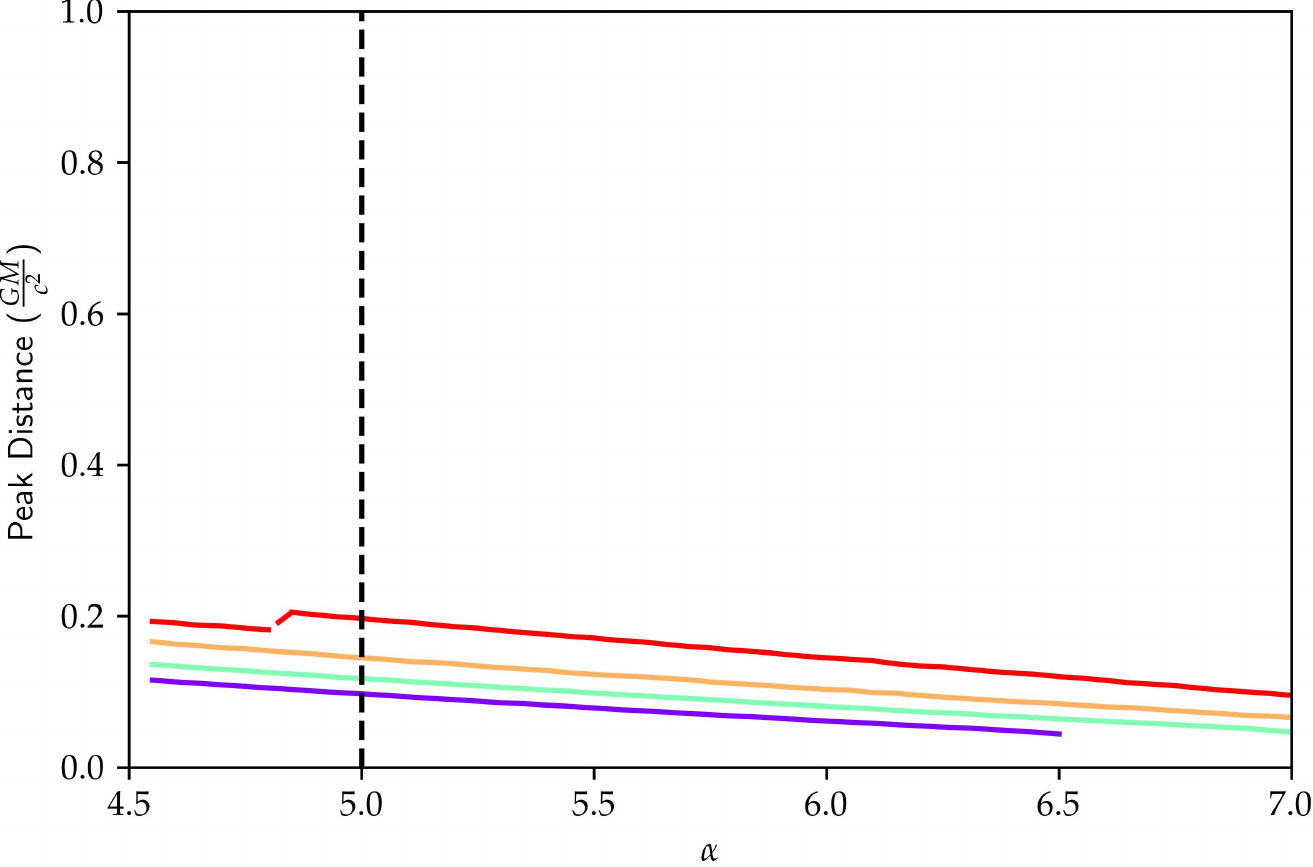}{.48\textwidth}{}
    \caption{Distance between the first and second intensity peaks at different spin with (top left) varying opening angle, $\theta_c$, (top right) observer inclination, $i$, (bottom left) observer axis, $\phi_i$ where $\tan(\phi_i)=\frac{y}{x}$, or (bottom right) $\alpha$, where $j_{\nu} \propto r^{-\alpha}$.  When not specified, inclination is fixed to $20\degree$, $\theta_c=60\degree$, $\phi_i=180\degree$, and $\alpha=5$.}
    \label{fig:peak_distance}
\end{figure*}

Figure \ref{fig:peak_distance} traces the distance between the first two intensity maxima against flow variables and observer inclinations. Discontinuities on the graph are indications that new intensity peaks have either formed or dissipated. The plot shows dependence on $\theta_c$, $i$, the orientation $\phi_c$ of the cut through the ring on the sky, and $\alpha$, the power-law index of the emissivity.   These are just four of the many variables that characterize black hole accretion flow, and have effects on the intensity peaks.  Therefore, while the definition of subrings are purely spacetime dependent, the intensity peaks derived from them are flow dependent.

\section{Conclusion} \label{sec:conclusion}

In images of black holes at millimeter wavelengths, some emission may be generated in the jet-disk boundary layer \citep{Wong_2021c}.  In this paper we have used a phenomenological dual-cone model to explore the implications of this possibility.  

Along the way we showed that, when viewed from outside the cone, the cone boundary is symmetric about the $y$ axis.  This implies that the jet centerline is not displaced perpendicular to the projection of the spin axis on the sky, while the center of the critical curve is displaced by $2 a \sin i$.  We have also computed images of the dual-cone model from outside and inside the cone.  In the former case the results are consistent with our analytic results and the jet centerline is unchanged by spin.  In the latter case an analytic result is not yet available, but we have shown numerically that the jet centerline is approximately unaffected by spin.  This suggests a method for measuring spin, by measuring the displacement of the ring center compared to the jet centerline.  Our results extend earlier work by \cite{Takahashi_2004} and \cite{Chael_2021}.   

We also analyzed the structure of brightness maxima near the critical curve of a black hole.  We find that the number and location of local maxima depends on the cone opening angle $\theta_c$.  For $\theta_c \simeq \pi/4$, for example, there are approximately two maxima for each subring.  Ring substructure is therefore flow-dependent.  If emission arises in a jet wall this implies that the ring substructure carries information about the jet and it may therefore be possible to constrain the jet opening angle close to the black hole. 

The dual cone model is only the lowest rung on a ladder of phenomenological models that might more accurately represent emission close to a black hole.  Imaging of GRMHD models is needed to explore the feasibility of using spin displacement to measure black hole spin.  

\section*{Acknowledgements}
This work is supported by NSF grants 17-16327, 17-43747, and 20-34306.  We thank the referee for comments that substantially improved the paper.
\bibliography{ref}{}
\bibliographystyle{aasjournal}
\end{document}